\def\la{\hbox{{\lower -2.5pt\hbox{$<$}}\hskip -8pt\raise
-2.5pt\hbox{$\sim$}}}
\def\ga{\hbox{{\lower -2.5pt\hbox{$>$}}\hskip -8pt\raise
-2.5pt\hbox{$\sim$}}}
\def\ltsima{$\; \buildrel < \over \sim \;$}
\def\simlt{\lower.5ex\hbox{\ltsima}}
\def\gtsima{$\; \buildrel > \over \sim \;$}
\def\simgt{\lower.5ex\hbox{\gtsima}}

\documentstyle[epsfig]{elsart}

\begin{document}
\begin{frontmatter}
\title{Space Time Fluctuations and Ultra High Energy Cosmic Ray
Interactions}
\author[lngs]{R. Aloisio},
\author[inaf]{P. Blasi},
\author[dipaq]{A. Galante},
\author[cnr,tor]{P.L. Ghia},
\author[lngs]{A.F. Grillo}
\address[lngs]{INFN - Laboratori Nazionali del Gran Sasso, SS. 17bis\\
Assergi (L'Aquila) - Italy}
\address[inaf]{INAF - Osservatorio Astrofisico di Arcetri, Largo E. Fermi 5\\
50125 Firenze - Italy}
\address[dipaq]{Dipartimento di Fisica, Universit\`a di L'Aquila, Via Vetoio\\
67100 Coppito (L'Aquila) - Italy}
\address[cnr] { CNR - IFSI, Sezione di Torino, Corso Fiume 4, 
10133 Torino - Italy}
\address[tor]{INFN - Sezione di Torino, 
Via P. Giuria 1, 10125 Torino - Italy}

\begin{abstract}
The intimate geometry of space-time is expected to suffer stochastic
fluctuations as a result of quantum gravitational effects. These 
fluctuations may induce observable consequences on the propagation of 
high energy particles over large distances, so that the strength and 
the characteristics of these fluctuations may be constrained, 
mainly in the range of energies of interest for cosmic 
ray physics. While invoked as a possible explanation for the detection 
of the puzzling cosmic rays with energies in excess of the threshold 
for photopion production (the so-called  super-GZK particles), we
demonstrate here that lower energy observations may provide 
strong constraints on the role of a fluctuating space-time structure.
\end{abstract}

\end{frontmatter}

\section{Introduction}

In the recent past it has been pointed out \cite{kir,lgm,cam,colgla,noi}
that observations in cosmic ray physics, and in particular Ultra High 
Energy Cosmic Ray (UHECR) experiments can be used to constrain to high
precision the possibility that the fundamental Lorentz invariance (LI) 
may be broken to some level.

In fact, violations of LI have been invoked by many authors 
\cite{cam,colgla,berto} to explain puzzling observations related to the
detection of cosmic rays with energy in excess of $\approx 10^{20}$ eV,
the so-called Greisen-Zatsepin-Kuzmin (GZK) \cite{gzk} cut-off, and the
more uncertain excess of photons in the TeV energy region from distant
sources. The two phenomena have in common that  in both cases 
the detected particles 
should instead suffer significant attenuation since their energy is above the  
threshold for photopion production (in the first
case) and pair production \cite{gamgam} (in the second case). In 
particular the GZK cutoff has been expected for a long time because of 
the dramatic energy losses of high energy protons with the $3^oK$ 
Cosmic Microwave Background Radiation (CMBR). Violations of LI in both
cases may act on the kinematics of the processes so to move the thresholds
in regions which are possibly outside the observational window. 
In the following we mainly concentrate upon the case of UHECRs, pointing
out, when necessary, the differences with respect to the case of pair 
production.

The recipes for the violations of LI that have been suggested insofar
generally consist of requiring an {\it explicit} modification of the 
dispersion relation of UHE particles, due to their propagation in 
the ``vacuum'', now affected by quantum gravity (QG). This effect is
generally parametrized by introducing a typical mass, expected to be of 
the order of the Plank mass ($M_P$), that sets the scale for QG to
become effective. 

In fact, explicit modifications of the dispersion relation are not really
necessary, as was recently pointed out in Refs. \cite{ford,ng1,ng2,lieu} 
for the case of propagation of UHECRs. There is an intrinsic uncertainty in the
process of measurement of physical quantities on time (space) scales 
comparable with the Planck time (length). It is generally believed that 
coordinate measurements cannot be performed with precision better than 
Planck distance (time) $\delta x \geq l_P$, since such a measurement 
would result in the production of a black hole. This means that the
metric of space-time must feature quantum fluctuations on the Planck scale.
A similar line of thought implies that an uncertainty in the measurement
of energy and momentum of particles must occur, as described by the  
relation $\delta p \simeq \delta E \simeq E^2/M_P$. According to the 
authors of Refs. \cite{ng1,ng2} the apparent problem of super-GZK particles
may find a solution also in the context of this uncertainty approach.

We discuss here this appealing approach more in detail, by taking into 
account the effects of the propagation of CRs in the QG vacuum in the 
presence of the universal microwave background radiation. A fluctuating
metric implies that different measurements of the particle energy or 
momentum may result in different outcomes. Therefore it becomes important to
define the probability that the {\it measured} energy (momentum) of a particle 
is above some fixed value. Note that averaging
over a large number of measurements would yield the {\it classical} values
for the energy and momentum. The process of measurement mentioned above, 
during the propagation of particles over cosmological distances occurs at
each single interaction of the particle with the environment. At each
interaction vertex, the fluctuating energy/momentum of the particle is 
compared with the kinematic threshold for the occurrence of some physical 
process (in our case the photopion production).
A clear consequence of this approach is that particles with classical energy
below the standard Lorentz invariant threshold have a certain probability of 
interacting. In the same way, particles above the classical threshold have a 
finite probability of evading interaction.

We show here that the most striking consequences of the approach described
above derive from low energy particles rather than from particles otherwise
above the threshold for photopion production. 
      
\section{The effect of Space-Time fluctuations on propagation
of UHE particles.}

While electroweak and strong interactions propagate through space-time, 
gravity turns out to be a property of the space-time itself. This simple 
statement has profound implications in that our belief that gravity can
be turned into a quantum theory immediately implies that the structure
of space-time has quantum fluctuations. Another way of rephrasing this 
concept is that space-time is expected to have a granular (or foamy) 
structure, where 
however the size of space-time cells fluctuates stochastically, thereby 
causing an intrinsic uncertainty in the measurements of space-time lengths, 
and indirectly of energy and momentum of a particle moving through space-time.
The uncertainty appears on scales comparable with the Planck scale.

It is generally argued that measurements of distances (times) smaller than 
the Planck length (time) are conceptually unfeasible, since the process of 
measurement collects in a Planck size cell an energy in excess of the Planck
mass, hence forming a black hole, in which information is lost.
This can be traslated in different ways 
into an uncertainty on energy-momentum measurements (\cite{ng1,ng2}). The
Planck length is a good estimate of the uncertainty in the De Broglie 
wave-length $\lambda$ of a particle with momentum $p$. Therefore 
$\delta \lambda \approx l_P$, and  $\delta p = \delta (1/\lambda) \approx 
(p^2 l_P)=(p^2/M_P)$.

Speculating on the exact characteristics of the fluctuations induced by
QG is beyond the scope of the present paper, and it would probably be
useless anyway, since the current status of QG approaches does not allow
such a kind of knowledge. We decided then to adopt a purely phenomenological
approach, in which some reasonable assumptions are made concerning the 
fluctuations in the fabric of space-time, and their consequences for the 
propagation of high energy particles are inferred.  Comparison with 
experimental data then possibly constrains QG models.

Following \cite{ng1}, we assume that in each measurement:

\begin{itemize}

\item{the values of energy (momentum) fluctuate around their average values
(assumed to be the result theoretically recoverable for an infinite number
of measurements of the same observable): 
\begin{equation}
 E \approx  {\bar E} + \alpha {{\bar E}^2\over M_P} 
\end{equation}
\begin{equation}
 p \approx  {\bar p} + \beta {{\bar p}^2\over M_P} 
\end{equation}
with $\alpha, \beta$ normally distributed variables and $p$ the modulus 
of the 3-momentum (for simplicity we assume rotationally invariant 
fluctuations);}

\item{the dispersion relation fluctuates as follows: 
\begin{equation}\label{eq:dis}
P_\mu g^{\mu\nu} P_\nu = E^2-p^2 + \gamma \frac{p^3}{M_P}=m^2
\end{equation}
and $\gamma$ is again a normally distributed variable.}

\end{itemize}

Ideally, QG should predict the type of fluctuations introduced above, but,
as already stressed, this is currently out of reach, therefore we assume here 
that the fluctuations
are gaussian. Our conclusions are however not sensitive to this assumption:
essentially any symmetrical 
distribution with variance $\approx 1$, within a large factor, would
give essentially the same results.
Furthermore we {\it assume} that  $\alpha$, $\beta$ and $\gamma$ 
are uncorrelated random variables; again, this assumption reflects our 
ignorance in the dynamics of QG 
\footnote{The fluctuations
described above will in general derive from metric fluctuations of
magnitude $\delta g^{\mu\nu} \sim h^{\mu\nu} \frac{l_P}{l}$ 
\cite{cam,ng2}. Our assumption reflects the fact that, while the magnitude of
the fluctuation can be guessed, we do not make any assumption on its 
tensorial structure $h^{\mu\nu}$.}. 

Our interest will be now concentrated upon processes of the type

$$a+b\to c+d$$

where we assume that a kinematic threshold is present; in the realm of 
UHECR physics (a,b) is either ($\gamma, \gamma_{3K}$)
or ($p,\gamma_{3K}$) and (c,d) is ($e^+,e^-$) or ($N,\pi$).

To find the value of initial momenta for which the reaction occurs we write
down energy-momentum conservation equations and solve them with the
help of the dispersion relations, as discussed in detail in \cite{noi}.

The energy momentum conservation relations are (in the laboratory frame,
and specializing to the case in which the target (b) is a low energy
background photon for which fluctuations can be entirely neglected)
\begin{equation}
E_a+ \alpha_a {E^2_a\over M_p} + \omega = 
E_c+ \alpha_c {E^2_c\over M_p}+ E_d + \alpha_d {E^2_d\over M_p}
\label{eq:disp1}
\end{equation}
\begin{equation}
p_a+ \beta_a {p^2_a\over M_p} - \omega = 
p_c+ \beta_c {p^2_c\over M_p}+ p_d + \beta_d {p^2_d\over M_p}.
\label{eq:disp2}
\end{equation}
These equations refer to head-on collisions and collinear
reaction products, which is appropriate for threshold computations. Together
with the modified dispersion relations, these equations, after 
some manipulations, lead to a cubic equation for the initial momentum 
as a function of the momentum of one of products, and, after minimization, 
they define the threshold for the process considered. In figure 1 we 
report the distribution of thresholds in the $\approx 70 \%$ of cases
in which the solution is physical; in the other cases the kinematics 
does not allow the reaction.

\begin{figure}[thb]
 \begin{center}
  \mbox{\epsfig{file=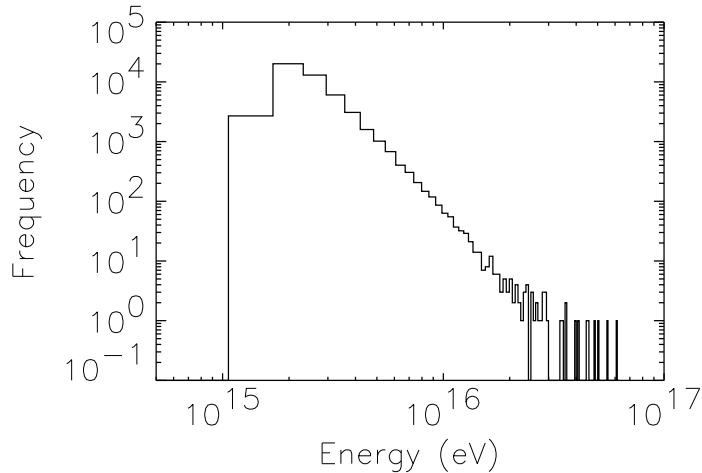,width=10.cm}}
  \caption{\em {Threshold distribution for $p \gamma_{3^oK} \to 
N \pi$. In the $30\%$ of cases the reaction is not allowed.}}
 \end{center}
\end{figure}
 This threshold distribution can be interpreted in the 
following way: a 
particle with energy above $\sim 10^{15}$ eV has essentially $70 \%$ 
probability of being above threshold, and therefore to be absorbed. In the
other $30 \%$ of the cases the protons do not interact.

In (4,5) the fluctuations are taken independently for each particle, which is 
justified as long as the energies are appreciably smaller than the Planck
energy. At that point it becomes plausible that different particles 
experience the same fluctuations, or more precisely fluctuations of the
same region of space-time. It is instructive to consider this case in some 
more detail: we introduce then the four-momenta (and dispersion 
relations) of {\it all} particles fluctuating in the same way. 
Specializing to proton interaction on CMBR, the equation which defines
the threshold $p_{th}$ is \cite{noi}:
\begin{equation}
\eta {2 p_0^3 \over { (m_{\pi}^2+2m_{\pi}m_p)  M_P}} 
{{m_{\pi}m_p} \over {(m_{\pi}+m_p)^2}} 
\left ( {p_{th} \over p_0} \right ) ^3   
+ \left ( {p_{th} \over p_0} \right )   -1 = 0
\end{equation}
where $\eta$ is a gaussian variable with zero average and variance
of the order of (but not exactly equal to) one, and $p_0$ is
the L.I. threshold (GZK). The threshold is the positive solution of
this equation.

The coefficient of the cubic term is very large, of the 
order of $10^{13}$ in this case, so that unless $\eta$ 
is $O(10^{-13})$,
we can write, neglecting pion mass
\begin{equation}
p_{th}\approx p_0 \left ( { m_p^2 M_P \over {\eta p_0^3 }} 
\right )^{1\over 3}.
\end{equation}
When $\eta$ becomes negative, the above equation has no 
positive root; this happens essentially in $50 \%$ of the cases.
Since the gaussian distribution is  flat in a small interval 
around zero, the distribution of thresholds for positive $\eta$ 
peaks around the 
value for $\eta \approx 1$, meaning that the threshold moves almost 
always down to a value of $\approx 10^{15}$ eV \cite{noi}; 
essentially the same result holds for fluctuations affecting only  the 
incident (higest energy) particle.
For {\it independent} fluctuations of final momenta,
the asymmetry in the probability of interaction arises from the fact that 
even exceedingly small negative values of the fluctuations lead 
to unphysical solutions. 

Building upon our findings, we now apply the same calculations to 
the case of UHECR protons propagating on cosmological distances.
An additional ingredient is needed to complete the dynamics of the
process of photopion production, namely the cross section. The rather
strong assumption adopted here is that the cross section remains the
same as the Lorentz invariant one, provided the reaction is 
{\it kinematically} allowed. This implies that the interaction lengths
remain unchanged.

In order to assess the situation of UHECRs, we first consider the 
case of particles above the threshold for photopion production in 
a Lorentz invariant world. According with eqs. (\ref{eq:disp1},
\ref{eq:disp2}),
in this case particles have a probability of $\approx 30 \%$ of being
not kinematically allowed to interact inelastically with a photon in the 
CMBR. Therefore, if our assumption on the invariance of the interaction 
length is correct, then each proton is still expected to make photopion
production, although with a slightly larger pathlength. 

The situation is however even more interesting for particles that are
below the Lorentz invariant threshold for the process of photopion 
production. If the energy
is below a few $10^{18}$ eV, a galactic origin seems to be in good 
agreement with measurements of the anisotropy of cosmic ray arrival 
directions \cite{agasa_a,fly_a}. We will not consider these energies 
any longer. On the other hand, at energies in excess of $10^{19}$ eV,
cosmic rays are believed to be extragalactic protons, mainly on the ground of
the comparison of the size of the magnetized region of our Galaxy and
the Larmor radius of these particles. 
We take these pieces of information as the basis 
for our line of thought. If the cosmic rays observed in the energy range 
$E> 10^{19} \rm{eV}$ are extragalactic protons,
then our previous calculations apply and we may expect that these particles 
have a $\sim 70\%$ probability of suffering photopion 
production, even if their energy is below the classical 
threshold for this process. Note that the pathlength 
associated with the process is of the order of the typical pathlength for 
photopion production (a few  tens of Mpc), therefore we are here discussing 
a dramatic process in which the absorption length of particles drops from 
Gpc, which would be pertinent to particles with energy below 
$\sim 10^{20}$ eV in a Lorentz invariant world, to several Mpc, with a 
corresponding suppression of the flux. 
 What are the consequences for the observed fluxes of cosmic rays? 
The above result implies that {\it all} with $E>10^{15}$ eV
are produced within a radius of 
several tens of Mpc, and above this energy there is no dramatic change 
of pathlength with energy. 
There is no longer anything
special about  $E \sim 10^{20}$ eV, and 
any mechanism invoked to explain the flux of super-GZK particles must
be at work also at lower energies. 

The basic situation remains the same in the case of pair production as the
physical process under consideration. For a source at cosmological distance,
a cutoff is expected due to pair production off the far infrared background
(FIR) or the microwave background. Using the results in \cite{noi} we expect
that the modified thresholds are a factor $0.06$ ($0.73$) lower than the 
Lorentz invariant ones for the case of interaction on the CMBR (FIR).
There is also a small increase in the pathlengths above the threshold,
which would appear exponentially in the expression for the flux. Therefore
there are two effects that go in opposite directions: the first moves the
threshold to even lower energies, and the second increases the flux of
radiation at Earth because of the increase of the pathlength. 
It seems that geometry fluctuations do not provide an immediate explanation
of the possible detection of particles in excess of the expected ones from
distance sources in the TeV region. In any case the experimental evidence 
for such an excess seems at present all but established.
 
\section{Discussion, conclusions and perspectives.}

The investigation of the dependence of the kinematic thresholds for 
physical processes has been shown to be a powerful tool to study the
Physics of space-time at extremely high energies, close to that Planck
scale where the fabric of space-time is expected to change, from 
the flat and well-behaved sheet that we experience in everyday life to 
a complicated and unpredictable foam of probability that, till now, 
evaded any kind of direct investigation. 

In this paper we related this foamy structure of space-time with the 
process of measurement of the physical properties of particles, in 
particular their energy and momentum. We found that the fluctuating 
metric may induce a violation of Lorentz invariance that changes the
thresholds for the photopion production of a very high energy proton 
off the photons of the CMBR, or for the pair production of a high energy
gamma ray in the bath of the FIR or CMBR photons.

For the case of UHECRs interacting with the CMBR, we obtained a picture that 
changes radically our view of the effect of QG on this phenomenon, as 
introduced in previous papers: not only particles with energy above 
$\sim 10^{20}$ eV are affected by the fluctuations in space-time, but also 
particles with lower energy, down to $\sim 10^{15}$ eV seem to be affected
by such fluctuations. In fact the latter, as a result of a fluctuating 
space-time, may end up being above the threshold for photopion production, 
so that particles may suffer significant absorption. Our conclusion is that 
all particles with energy in excess of $\sim 10^{15}$ eV eventually detected 
at Earth would be generated at distances comparable with the pathlength 
for photopion production ($\sim 100$ Mpc). 
A consequence of this is that there is no longer anything special 
characterizing the energy $\sim 10^{20}$ eV. 

Since the conclusion of our work are quite strong, it is important to 
summarize in detail the assumptions involved in the calculations:

1) space-time fluctuations follow from Quantum Gravity. This is a rather 
mild statement and essentially accepted on the basis of the ``space-time foam''
approach \cite{wheeler57}. 
Following \cite{ng1,ng2} we further assume that these fluctuations are 
at the level of the Planck scale \footnote{We assume that this 
is true in the comoving reference frame, in which the CMB radiation is 
isotropic. If this is assumed to be true in {\it all} reference 
frames, than it is necessary to modify the Lorentz transformations
\cite{amelino,smolin}}. 

2) Fluctuations in the momentum and in the dispersion relation are 
induced by the fluctuations in the metric of space-time. Our assumption
is that the fluctuations in the dispersion relation and in the energy and
momenta of the particles involved can be considered as independent. This
is an assumption usually shared by most literature in the field.

3) The cross section for photopion production at threshold is assumed to be 
the same as in the Lorentz invariant case. The cross section can be considered
as a combination of a matrix element and a phase space factor. While the 
former may well be affected by violation of Lorentz invariance, it can be 
demonstrated that the phase space does not change appreciably once the
incident particle has energy above the threshold for photopion production.
This may be not true in other situations, as discussed in the literature
\cite{amepai,stanev}.

We list below some tests that may allow to understand whether the
current or future observations are compatible with the scenario discussed
in this paper.

a)  Future experiments \cite{auger} 
dedicated to the detection of UHECRs will provide a
substantial increase in the statistics, so that the spectral features of 
the UHECRs in the energy region $E>10^{19}$ eV can be resolved, and
 further indications on the nature of primaries and their 
possible extragalactic origin will be obtained. In particular
the present possible disgreement between  
AGASA \cite{agasa_f} and HiRes \cite{hires} will be clarified.

One should also keep in mind that an evaluation of the expected flux in 
terms of sources distributed as normal galaxies is
in contradiction with AGASA data by an amount ranging from $2$ to $6 \sigma$
depending on the assumed source spectrum \cite{blanton}. 
Since the nature of the sources is not known, it is not
clear if their  abundance within the absorption pathlength
is sufficient to explain the observed flux in presence of space-time 
fluctuations, nor if they can induce observable anisotropies.

In any case, in a Lorentz invariant framework a suppression in the flux 
at $\sim 10^{20}$ eV is expected. If such a feature is 
unambiguously detected in the UHECR spectrum, no much room would be left for
the fluctuations of space-time discussed in this paper, since in this scenario
nothing special happens around $10^{20}$ eV.
In quantitative terms \cite{noi} 
this would imply  a phenomenological bound on $\l_P$ 
now interpreted as a parameter: 
 $l_p < 10^{-46}$ cm instead of 
$l_p \approx 10^{-33}$ cm; in other words, only fluctuations with 
variance $\approx 10^{-13}$, instead of $1$, would be allowed
\footnote{ Alternatively, 
one can assume a more general form of fluctuations, i.e. $\delta E \approx 
E (E/M_P)^{\alpha}$ and similar for momentum and dispersion relations. In 
this case a bound $\alpha > 2.3$ would follow.}.

b) According with our findings, all particles with energy in excess of 
$\sim 10^{15}$ eV lose their energy by photopion production on cosmological
spatial scales, as a result of the metric fluctuations. This energy ends up 
mainly in gamma rays, neutrinos and protons. The protons pile up in the
energy region right below $\sim 10^{15}$ eV. The gamma ray component
actually generates an electromagnetic cascade that ends up contributing 
low energy gamma rays, in the energy band accessible to instruments like
EGRET and GLAST. This cascade flux cannot be larger than the measured 
electromagnetic energy density in the same band
$\omega_{cas}^{exp}=10^{-6}~eV/cm^{3}$ \cite{egret}. The cascade flux in our 
scenario can be estimated as follows. Let $\Phi(E)=\Phi_0 (E/E_0)^{-\gamma}$
be the emissivity in UHECRs ($\rm{particles}/cm^3/s/GeV$). Let us choose the
energy $E_0=10^{10}$ GeV and let us normalize the flux to the observations
at the energy $E_0$. The total energy going into the cascade can be shown 
to be $$\omega_{cas} \approx \frac{5\times 10^{-4}}{\gamma-2}~~ 
x_{min}^{2-\gamma} ~~\xi ~~eV ~cm^{-3},$$ 
where $\xi$ is the fraction of energy going into gamma rays in each 
photopion production, and $x_{min}=(E_{th}/E_0)=10^{-4}$ for $E_{th}=
10^{15}$ eV. It is easy to see that, for $\gamma=2.7$, the cascade bound 
 is violated unless 
$\xi \ll 10^{-3}$.

One note of warning has to be sent concerning the development of the 
electromagnetic cascade: the same violations of LI discussed here affect 
other processes, as stressed in the paper. For instance pair production 
and pion decay are also affected by violations of LI \cite{amepai}. 
Therefore the possibility that the cascade limit is exceeded concerns 
only those scenarios of violations of LI that do not inhibit appreciably 
pair production and the decay of neutral pions.

The protons piled up at energies right below $10^{15}$ eV, would be a 
nice signature of this scenario, but it seems difficult to envision a 
way of detecting these remnants. In fact, even a tiny magnetic field
on cosmological scales would make the arrival time of these particles to
Earth larger than the age of the universe. Moreover, even assuming an exactly
zero extragalactic magnetic field, these particles need to penetrate the 
magnetic field of our own Galaxy and mix with the galactic cosmic rays, 
making their detection extremely problematic if not impossible.

We believe that the results presented here are robust, in the sense
that the qualitative features are rather insensitive to modifications
of the theoretical scheme. Clearly a more detailed flux computation,
taking into account propagation of primaries as well as generation and
propagation of the secondaries is needed in order to assess in a more
quantitative way observable effects of possible metric fluctuations on UHECRs.
 
{\bf Acknowledgements} We are grateful to Giuseppe Di Carlo for very useful 
and stimulating discussions.

\begin{small}
\begin{center}
{\bf Note Added}
\end{center}
Soon after the completion of the present work,
a paper \cite{lastcamel} appeared in the preprint archives which
addresses the problem of the fluctuations in the metric of space-time
and the corresponding violations of LI. Our results include the results
of that paper, but also represents a generalization to the 
particles with energy below the classical threshold for 
photopion production, that in our opinion imply a possibly stronger 
constrain on the effect of fluctuations of space-time on cosmic rays
propagation.
\end{small}


\begin{thebibliography}{99}

\bibitem{kir}
D.A. Kirzhnits and V.A. Chechin, Sov. Jour. Nucl. Phys. {\bf 15}, 585 (1971).

\bibitem{lgm}
L. Gonzalez-Mestres, Proc. 26th ICRC (Salt Lake City, USA), {\bf 1}, 179 
(1999).

\bibitem{cam}
G. Amelino Camelia, J. Ellis, N.E. Mavromatos and S. Sarkar
Nature {\bf 393} (1998) 763.

\bibitem{colgla}
S. Coleman, S.L. Glashow, Phys. Rev. {\bf D59}, 116008 (1999).

\bibitem{berto}
O. Bertolami and C.S. Carvalho, Phys. Rev. {\bf D61}, 103002 (2000);
O. Bertolami, {\it preprint} astro-ph/0012462.
 
\bibitem{noi}
R. Aloisio, P. Blasi, P.L. Ghia and A.F. Grillo,
Phys. Rev. {\bf D62}, 053010 (2000).

\bibitem{gzk}
K. Greisen, Phys. Rev. Lett. {\bf 16}, 748 (1966);
G.T. Zatsepin and V.A. Kuzmin, Pis'ma Zh. Ekps. Teor. Fiz. {\bf 4}, 114  
(1966) [JETP Lett. {\bf 4}, 78 (1966)].

\bibitem{gamgam}
A.I. Nikishov, Sov. Phys. - JETP {\bf 14}, 393 (1962);
P. Goldreich and P. Morrison, Sov. Phys. - JETP {\bf 18}, 239 (1964);
R.J. Gould and G.P. Schreder, Phys. Rev. Lett. {\bf 16}, 252 (1966).

\bibitem{ford}
L.H. Ford Int. J. Theor. Phys. {\bf 38} 2941 (1999).

\bibitem{ng1}
Y.J. Ng, D.S. Lee, M.C. Oh and H. van Dam, 
Phys. Lett. {\bf B507} 236 (2001).

\bibitem{ng2}
Y.J. Ng, {\it preprint} astro-ph/0201022.

\bibitem{lieu}
R. Lieu, Astrophys. J. {\bf 568} L67 (2002).

\bibitem{agasa_a}
N.Hagashida et al., Astrop. Phys. {\bf 10} 303 (1999).

\bibitem{fly_a} 
D.J. Bird et al. Ap. J. {\bf 511} 739 (1999).

\bibitem{wheeler57}
J.A. Wheeler, Annals of Physics {\bf 2}, 604 (1957).

\bibitem{amelino}
G. Amelino-Camelia, {\it preprint} gr-qc/0012051, hep-th/0012238

\bibitem{smolin}
J. Maguejo and L. Smolin, {\it preprint} hep-th/0112090

\bibitem{amepai}
G. Amelino-Camelia, Phys. Lett. {\bf B528} (2002) 181.

\bibitem{stanev}
H. Vankov and T. Stanev, {\it preprint} astro-ph/0202388.

\bibitem{auger} J. W. Cronin, Nucl. Phys. B Proc. Suppl. {\bf 28B}, 213
(1992).

\bibitem{agasa_f} 
N. Sakay et al. Proceedings of 2001 ICRC.

\bibitem{hires}
C.C.H. Jui et al. Proceedings of 2001 ICRC.

\bibitem{blanton}
M. Blanton, P. Blasi and A.V. Olinto, Astrop. Phys., 15, 275 (2001).

\bibitem{egret}
P. Sreekumar et al. (EGRET collaboration), Astroph. J. {\bf 494} (1998) 523.

\bibitem{lastcamel}
G. Amelino-Camelia, Y.J. Ng and H. Van Dam, {\it preprint} gr-qc/0204077.

\end{thebibliography}
\end{document}